\documentclass[letter]{aa}

\usepackage{graphicx}
\usepackage{txfonts}
\usepackage{booktabs}
\usepackage[normalem]{ulem}
\usepackage{multirow}
\usepackage{hyperref}
\usepackage{orcidlink}
\newcommand{\newgalaxy}{UVIT J2022}
\begin{document} 

   \title{Hidden in Plain Sight: UVIT and MUSE discovery of a Large, Diffuse Star-Forming Galaxy}


   \author{Jyoti Yadav \orcidlink{0000-0002-5641-8102}
          \inst{1,}\inst{2}\thanks{jyoti [at] iiap.res.in}
          \and
          Mousumi Das \orcidlink{0000-0001-8996-6474} \inst{1}
          \and
          Sudhanshu Barway \orcidlink{0000-0002-3927-5402} \inst{1}
          \and
          Francoise Combes \orcidlink{0000-0003-2658-7893} \inst{3}}

   \institute{Indian Institute of Astrophysics, Koramangala II Block, Bangalore 560034, India\\
  \and
Pondicherry University, R.V. Nagar, Kalapet, 605014, Puducherry, India\\
\and
Observatoire de Paris, LERMA, College de France, CNRS, PSL University, Sorbonne University, F-75014 Paris, France}


 
  \abstract
{We report the discovery of a nearby large, diffuse galaxy that shows star formation, using Ultra Violet Imaging Telescope (UVIT) far-UV observations, archival optical data from Multi-Unit Spectroscopic Explorer (MUSE) and Dark Energy Camera Legacy Survey (DECaLS), and InfraRed Survey Facility (IRSF) near-infrared observations. The galaxy was not detected earlier due to its superposition with the background galaxy, NGC\,6902A. They were together mistakenly classified as an interacting system. NGC\,6902A is at a redshift of 0.05554, but MUSE observations indicate that the interacting tail is a separate star-forming, foreground galaxy at a redshift of 0.00980. We refer to the new galaxy as UVIT\, J202258.73-441623.8 (\newgalaxy).
The near-infrared observations show that \newgalaxy\ has a stellar mass of 8.7$\times$10$^{8}$M\textsubscript{\(\odot\)}. Its inner disk (R$<$4 kpc) shows UV and H$\alpha$ emission from ongoing massive star formation. The rest of the disk is extremely low luminosity, has a low stellar surface density, and extends out to a radius of R$\sim$9 kpc. The velocity and metallicity distribution maps and the star formation history indicate that \newgalaxy\ has undergone three bursts of star formation. The latest episode is ongoing, which is supported by the presence of widespread H$\alpha$ and UV emission in its inner disk. The galaxy also shows patchy spiral arms in far-UV, and there is a metallicity enhancement along a bar-like feature. \newgalaxy\ is thus a unique example of triggered star formation in a diffuse galaxy, resulting in the growth of its inner stellar disk. Our study raises the intriguing possibility that (i) there could be similar diffuse galaxies that have been mistakenly interpreted as interacting galaxies due to their superposition, and (ii) UV or H$\alpha$ could be a way to detect such diffuse galaxies in our local universe.
}
   \keywords{Galaxies: individual: NGC\,6902A, UVIT\,J202258.73-441623.8 -- Galaxies: interactions -- Galaxies: star formation -- Techniques: imaging spectroscopy
  }

   \maketitle
%

\section{Introduction} \label{sec:intro}
Low surface brightness galaxies (LSBGs) are defined as galaxies with very diffuse stellar disks having central surface brightness in B band ($\mu_{B}$) fainter than 23~mag/arcsec$^2$ \citep{impey.bothun.1997}. However, in the past decade, several deep optical surveys have detected LSBGs down to much lower faintness levels of 26~mag/arcsec$^2$ in the r band \citep{venhola.etal.2018,lim.etal.2020}.
The extremely faint and extended LSBGs are called ultra-diffuse galaxies (UDGs). Although the criteria vary in the literature, in general, UDGs have a central g band ($\mu_{g}(0))$ surface brightness fainter than 24~mag/arcsec$^2$ and an effective radius larger than 1.5~kpc \citep{vandokkum.etal.2015}. UDGs are classified as blue (gas-rich) and red (gas-poor) UDGs. The gas-rich UDGs are present in the outskirts of galaxy groups. In contrast, the gas-poor UDGs are found in clusters, which suggests that the environmental processes
 are dominating \citep{leisman.etal.2017, Prole2019MNRAS.488.2143P}.
These galaxies make up the tail end of the luminosity function \citep{koda.etal.2015}, and may extend down to much lower levels than we expect \citep{fattahi.etal.2020}. The common unifying factor for all LSBGs and UDGs is that they have diffuse stellar disks, i.e. their disks appear to have low stellar surface densities \citep{sales.etal.2020}, which can be connected to low star formation rates \citep{rong.etal.2020}. Although LSBGs are historically considered to be gas-rich \citep{honey.etal.2018}, UDGs may or may not contain gas \citep{leisman.etal.2017,chowdhury.etal.2019}.

Another class of galaxies that have very low stellar density disks are the extended UV (XUV) disk galaxies, which show star formation in their extreme outer disks, where both stellar densities and metallicities are very low \citep{Yadav2021ApJ...914...54Y}. The star formation in these adverse environments may have been triggered by gas inflow from nearby galaxies or cold gas accretion from the cosmic web \citep{sancisi.etal.2008}. Taking this as
an example, it is possible that UV or H$\alpha$ emission from compact star-forming complexes (SFCs) in diffuse galaxies could be one of the ways to detect LSBGs or blue UDGs that may otherwise go unnoticed in larger surveys. UV emission is especially important because it traces young O, B stars for $10^8$ years, whereas H$\alpha$ emission is produced by photoionization from massive stars (M$>$10M\textsubscript{\(\odot\)}) and is sensitive to star formation for only $10^6$ to $10^7$ years \citep{das.etal.2021}. Star formation is, however, difficult in low-density disks as the disk self-gravity is low. So triggers such as galaxy interactions and flybys are important, as shown in semi-analytical simulations \citep{Somerville2001MNRAS.320..504S, Menci2005ApJ...632...49M}. 
This letter reports the serendipitous discovery of a nearby diffuse galaxy showing intense star formation in its inner disk using  Ultra Violet Imaging Telescope (UVIT) and Multi-Unit Spectroscopic Explorer (MUSE) data. It is part of a larger study of star formation in a sample of southern interacting galaxies. This new galaxy lies in the foreground of NGC\,6902A, which, as shown in this paper, has been mistakenly classified as an interacting galaxy. The following sections describe our results and their implications for detecting LSBGs in our nearby universe. We used flat cosmology with $\Omega_{\Lambda}$=0.7, $\Omega_{M}$=0.3, H$_{0}$= 70 km s$^{-1}$ Mpc$^{-1}$ in this paper.

\begin{figure*}
   \begin{centering}
    \includegraphics[width=0.9\textwidth]{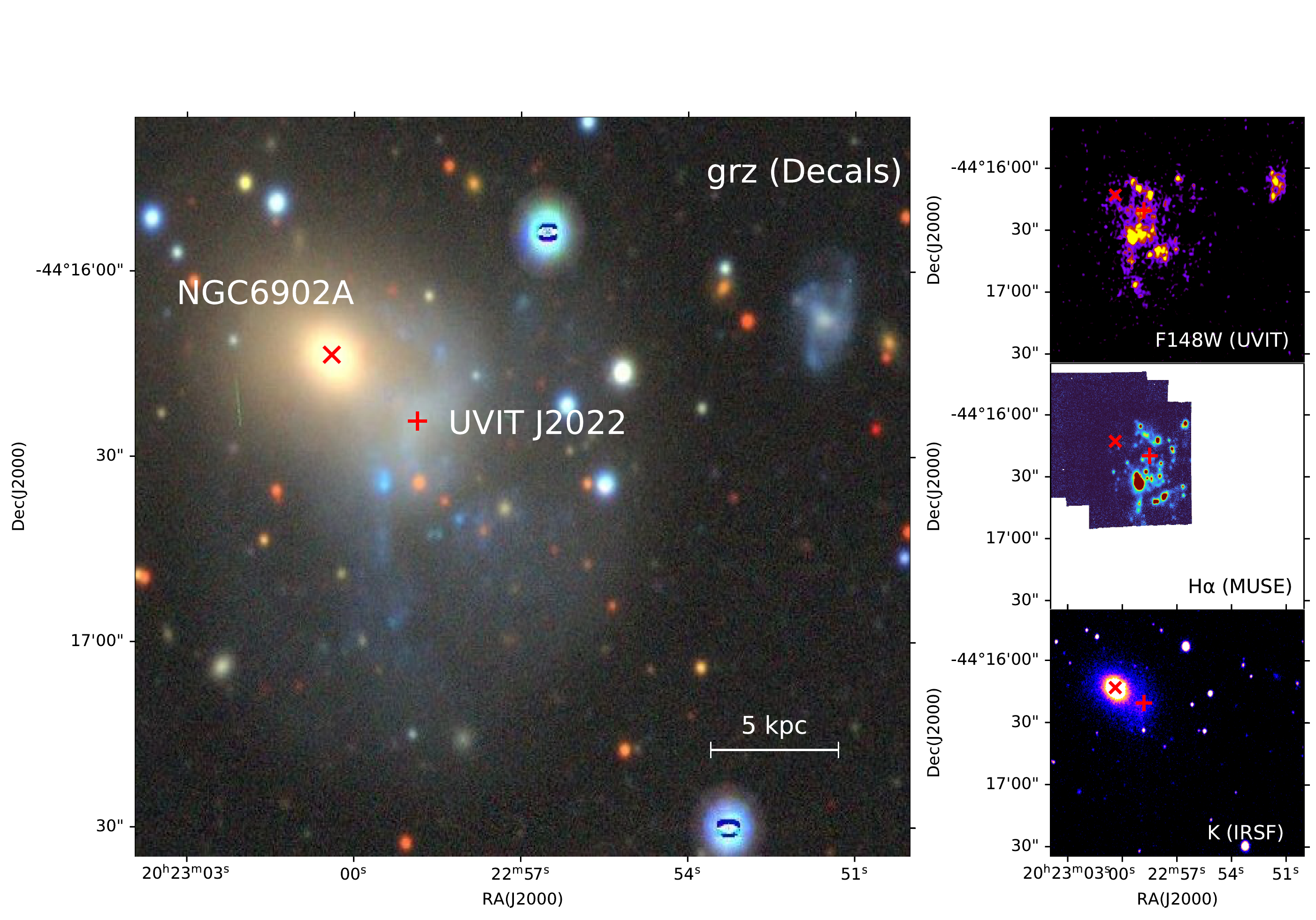}
    \end{centering}
    \caption{Multiwavelength images of \newgalaxy. 
    The grz image shows the background galaxy NGC\,6902A and foreground galaxy \newgalaxy. Star-forming regions in \newgalaxy\ are prominent in FUV and H$\alpha$. The K band image reveals the emission from old stars in NGC\,6902A. The $+$ and x symbol in red indicates the kinematic centre of the \newgalaxy\ and NGC\,6902A.} 
    \label{fig:images}
\end{figure*}

\section{Observations}
\subsection{Imaging Data}
We observed NGC\,6902A using the UVIT on board \textit{Astrosat} \citep{2012Kumar}. The UVIT has two telescopes, one for far-ultraviolet (FUV; 1300--1800 \AA), another for near-ultraviolet (NUV; 2000--3000 \AA). The UVIT has a field of view of 28$\arcmin$ and spatial resolution of $\sim$1\arcsec. The FUV observations were done in the F148W band (1231--1731 \AA). We reduced the UVIT level 1 data using \textsc{ccdlab} \citep{2017PASP..129k5002P,2020PASP..132e4503P}. 

We used archival g, r and z band data from the Dark Energy Camera Legacy Survey (DECaLS). DECaLS uses the Dark Energy Camera (DECam; \citealt{Flaugher2015AJ....150..150F}) mounted on the Victor M. Blanco telescope at the Cerro Tololo Inter-American Observatory (CTIO). DECam has 62 CCDs with 2048 × 4096 pixel format each for imaging. It has a field of view of 2.2$^{\circ}$ in diameter and a pixel scale of 0.262 arcsec/pixel. DECam can achieve 5$\sigma$ depth in total exposure times of 166, 134 and 200 sec in g, r, and z band for an emission line galaxy with a half-light radius of 0.45$\arcsec$.

The near-infrared observations of NGC\,6902A were carried out using SIRIUS camera \citep{2003SPIE.4841..459N} on the Infrared Survey Facility (IRSF) 1.4 m telescope at South African Astronomical Observatory (SAAO) Sutherland, South Africa. The SIRIUS camera performs simultaneous imaging in the \textit{JHK} bands. It has a field of view of 7.7\arcmin\ $\times$ 7.7\arcmin \citep{2012Nagayama}. The exposure time of the images is $\sim$120 minutes, and they were taken in automatic dithering mode with $\sim$20\arcsec\ steps with individual frame exposure times of 30 seconds each. We used the pipeline available for the SIRIUS observations to reduce the data, including corrections for nonlinearity, dark subtraction, and flat fielding. 
\subsection{Integral Field Unit Data}\label{sec:IFU}

We used MUSE archival data \citep{Bacon2010SPIE.7735E..08B}. MUSE gives 3D spectroscopic data cubes with high resolution. We used the data from the wide-field mode, which has a field of view of 1\arcmin\ $\times$ 1\arcmin\ and a spectral resolution of 1750 at 4650{\AA} to 3750 at 9300\AA. We used the Galaxy IFU Spectroscopy Tool (\textsc{gist}\footnote{\url{https://abittner.gitlab.io/thegistpipeline/}} version 3.0.3; \citealt{2019Bittner}) pipeline to study the properties of \newgalaxy. \textsc{gist} uses python implemented version of GANDALF \citep{2006Sarzi, 2006Falcon, 2019Bittner} and Penalized Pixel-Fitting, \citep[pPXF;][]{2004Cappellari,2017Cappellari} to provide emission-line properties and stellar kinematics, respectively. We Voronoi binned the data based on H$\alpha$ (6558--6568 \AA) emission. We used a signal-to-noise (S/N) of 30 for binning the data. We fitted the continuum using a multiplicative eighth-order Legendre polynomial. We removed noisy data below S/N of 5 before binning the data. 

\section{Results} \label{sec:results}
The left panel of Fig.~\ref{fig:images} shows the DECaLS \textit{grz} color image for NGC\,6902A which is classified as an interacting galaxy \citep{1991deVaucouleurs}. The south-west outer region of NGC\,6902A in the DECaLS colour image shows diffuse blue emission. This southwestern region shows prominent SFCs in the FUV image and corresponds to strong H$\alpha$ emission from the MUSE cube as shown in the right panel of Fig.~\ref{fig:images}.
NGC\,6902A is bright in the K band image, and the disk appears extended in the south-west. However, there is only faint FUV emission and almost no H$\alpha$ emission from NGC\,6902A at the rest-frame wavelength. 

This prompted us to investigate the peculiar feature in more detail to determine the cause of interactions to understand if this feature is the remnant of another galaxy gone through a merger with NGC\,6902A. To do so we used H$\beta$, [\ion{O}{iii}], H$\alpha$, and [\ion{N}{ii}] lines of various SFCs in this region for redshift estimation. The mean redshift calculated from the emission lines for these SFCs is z=0.00980$\pm$ 0.00018, whereas the redshift of NGC\,6902A is z=0.05554 \citep{Costa1991ApJS...75..935D}. It means that the diffuse blue emission was, in fact, from a foreground galaxy that we discovered using FUV, MUSE and K band data. We have named it UVIT\, J202258.73-441623.8 (hereafter referred to as \newgalaxy). In the following sections, we described the detailed study to establish the characteristics and morphology of \newgalaxy.

\begin{table}
    \centering
    \begin{tabular}{cc}
    \toprule
    Source & UVIT\, J202258.73-441623.8\\
    \hline
    R.A (J2000)& 20:22:58.73 \\
    Dec. (J2000)& -44:16:23.8  \\
    z  & 0.00980$\pm$ 0.00018 \\
    Distance  & $\sim$ 41.86 Mpc \\
    V$_{sys}$ & $\sim$ 2930 kms$^{-1}$ \\
    Inner disk radius & $\sim$ 4 kpc \\
    Outer disk radius & $\sim$ 9 kpc \\
    HI mass  & 1.32$\times$10$^{9}$ M$_{\odot}$ \\
    Stellar Mass &{8.72$\times$10$^{8}$ M$_{\odot}$} \\
   { log($\Sigma_{SFR}$(M$_\odot$ yr$^{-1}$ kpc$^{-2}$))} FUV &  -0.74 \\
    {log($\Sigma_{SFR}$(M$_\odot$ yr$^{-1}$ kpc$^{-2}$))} H$\alpha$ &  -0.68\\
    \toprule
    \end{tabular}
    \caption{Details of \newgalaxy. The HI mass excludes the contribution of helium}.
    \label{tab:details}
\end{table}

\subsection{Neutral Gas Content, Gas Kinematics and Metallicity } 
 To determine the gas content of \newgalaxy\ we looked for neutral gas line emission (HI) in the HIPASS database, using the location of \newgalaxy\ and the appropriate redshift. We found some weak emission associated with the galaxy at an approximate velocity of $v_{sys}$=2930 kms$^{-1}$. The flux value is $\sim$0.032 Jy/beam. Since the HIPASS beam is quite large ($\sim15.5^{\prime}$) and the diameter of \newgalaxy\ is $\sim1.5^{\prime}$, we can safely assume that all the HI emission lies within the beam. The neutral gas mass including contribution from HI gas and helium is given by, $M =1.4\times 2.36\times10^{5}\times D_{Mpc}^{2} \times S_{\nu}\Delta V$ = $1.85\times {10}^{9} M_{\odot}$, where $D_{Mpc}$ is the distance in Mpc to \newgalaxy, $S_{\nu}$ is the flux in Jy and $\Delta V$ is the width of the line in kms$^{-1}$ and $\Delta V \sim 100$kms$^{-1}$. Excluding the contribution of helium, the HI gas mass is $M(HI)=1.32\times10^{9} M_{\odot}$.

\begin{figure}
    \centering
    \includegraphics[width=0.49\textwidth]{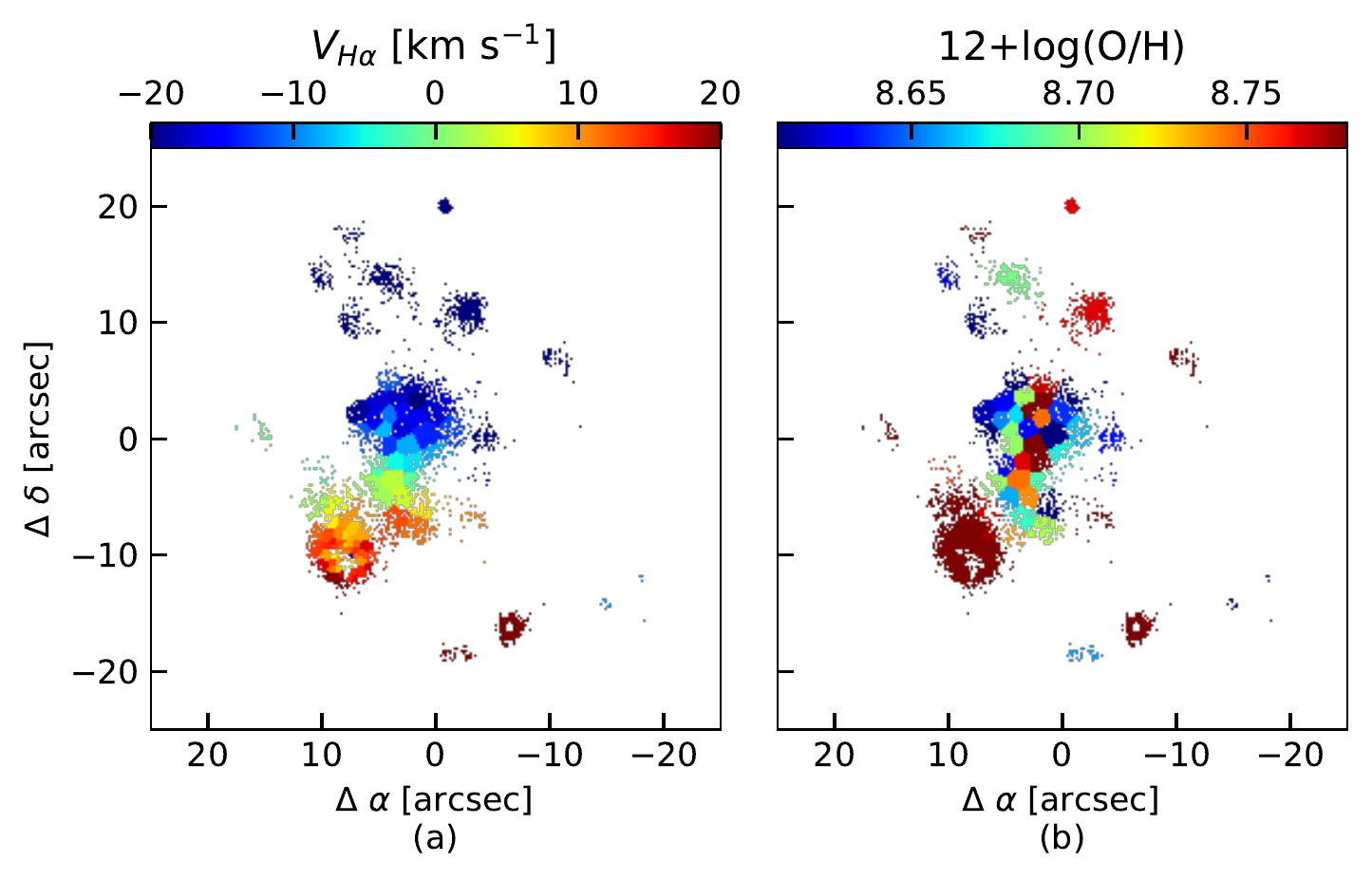}
    \caption{Panel(a) and panel (b) shows the H$\alpha$ velocity map and metallicity map, respectively. The maps show the bins with an amplitude-to-noise ratio of more than 4 in H$\beta$, [\ion{O}{iii}], H$\alpha$, and [\ion{N}{ii}] lines. Origin (0,0) in the maps is the coordinates of Table~\ref{tab:details}.}
    \label{fig:vel_metall_sfh}
\end{figure}

\newgalaxy\ is much brighter in H$\alpha$ than the old stellar disk, which is faint and diffuse in the IRSF K band image. We used H$\alpha$ to derive the velocity field of this galaxy. 
Fig.~\ref{fig:vel_metall_sfh}(a) shows the H$\alpha$ velocity map of the galaxy. It indicates galactic rotation with a velocity range of $\sim$ -20 to 20 km s$^{-1}$. The smaller range in velocity could be due to the face-on orientation of the galaxy.

Fig.~\ref{fig:vel_metall_sfh}(b) shows the metallicity map of \newgalaxy\ derived using \citet{Pettini2004MNRAS.348L..59P}.
\begin{equation}\label{eq:metallicity}
    12+log(O/H)= 8.73-0.32\times log[([\ion{O}{iii}]/H\beta)/ ([\ion{N}{ii}]/H\alpha)]
\end{equation}
The metallicity map shows that the galaxy has higher metallicity along the northwest to southeast direction, which could be due to a small bar-like feature in the stellar disk of \newgalaxy. The SFCs at the edges of the bar also show higher metallicity. Bars are known to funnel gas from the outer to the inner disks, leading to star formation, thus enhancing the inner disk metallicities of galaxies \citep{combes.etal.2014}. The F148W and H$\alpha$ images also show bright SFCs along and at the edges of the bar.

\begin{figure}
    \centering
    \includegraphics[width=0.5\textwidth]{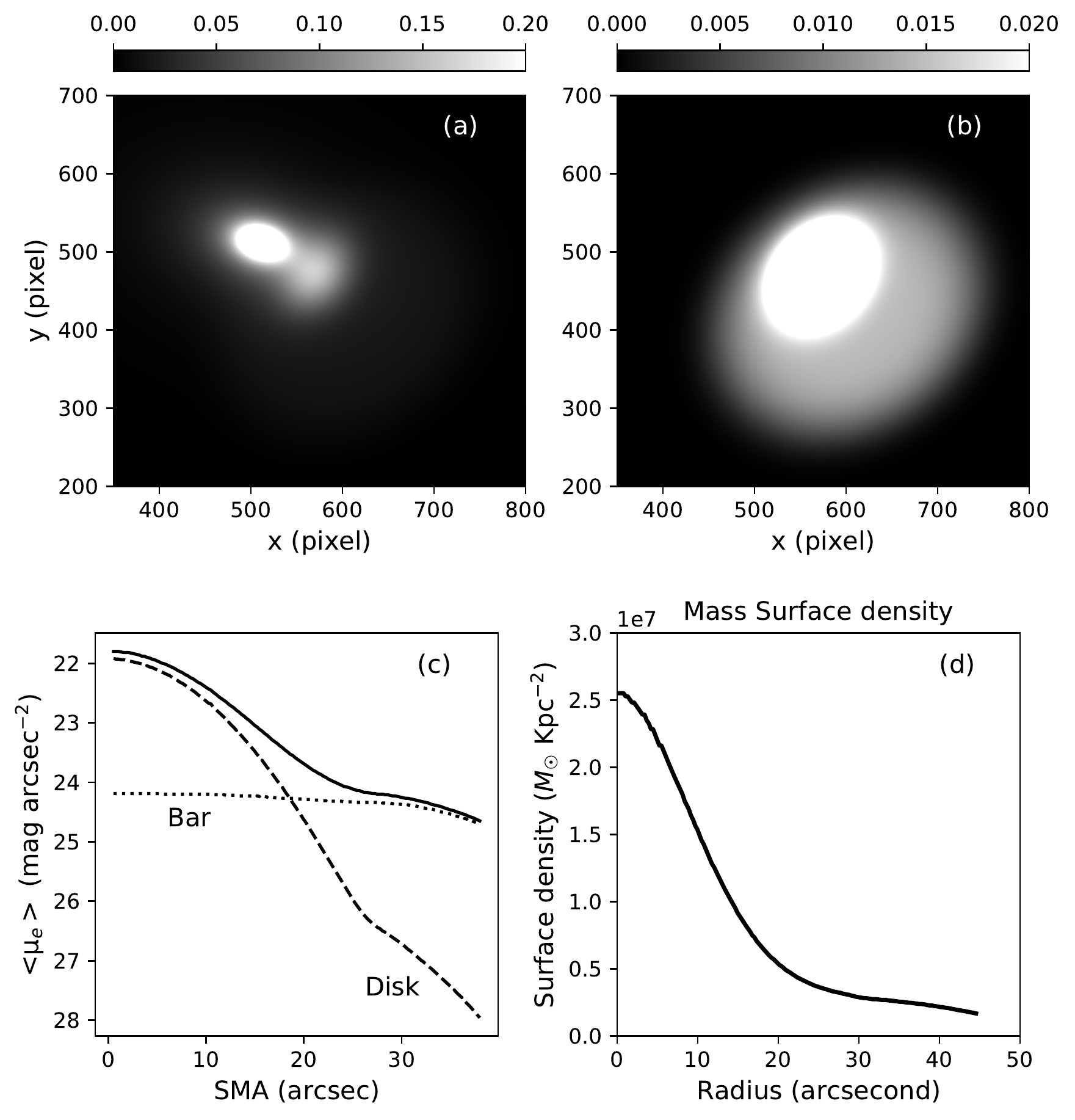}
    \caption{Panel (a) shows the model obtained from fitting of NGC\,6902A and \newgalaxy\ with 3+2 Sersic component in {r} band. Panel (b) shows the model image of \newgalaxy. The size scale is the same in panel (a) and (b). However, the grayscale in panel (b) is different to show the extent of \newgalaxy. Panel (c) shows the total model{ average} surface brightness profile of \newgalaxy\ in {r} band, obtained from \textsc{galfit} (solid curve). The dotted and dashed curves show the Sersic profiles for the inner central bar like structure and disk component, respectively. Panel (d) shows the surface density of \newgalaxy.}
    \label{fig:brightness_profile}
\end{figure}

\subsection{Decomposition using GALFIT} 
We used \textsc{galfit} \citep{2002AJ....124..266P} to perform the 2-D decomposition of NGC 6902A and \newgalaxy\ in the {r} band. Although we did have IRSF K band data, the diffuse nature of the stellar disk made it difficult to quantify the disk properties with good S/N; hence we used the {r} band image instead. A good fit was obtained by fitting three Sersic functions to NGC 6902A and two Sersic functions to the foreground galaxy \newgalaxy. The best fit was used to obtain models for both galaxies and are shown in Fig.~\ref{fig:brightness_profile}. We found that the effective radius, $r_e$, of the first Sersic profile in the {r} band is {7.57} kpc which probably corresponds to a large diffuse disk. The effective radius for the second Sersic profile in the {r} band is {2.35} kpc, corresponding to a central stellar bar. However, the bar is not visible in the K band image, which may be due to its diffuse nature. The central disk brightness for the outer diffuse disk in the r band is $\mu_{0}$(r)=20.13 mag arcsec$^{-2}$, after correcting for inclination and cosmological dimming. {However, there can be contamination due to background galaxy NGC\,6902A.}

To understand the stellar disk in more detail, we used the surface brightness profile, i.e. the magnitude/arcsec$^2$ in the z band, to derive the mass density. The mass to light ratio assumed for the z band was $M/L=1.4$ \citep{du.mcgaugh.2020}, which gives a total stellar mass of M(*)=$8.72\times10^{8}M_{\odot}$. Using this stellar mass M(*) in the mass-metallicity relation \citep{Tremonti2004ApJ...613..898T}, the metallicity of UVIT J2022 is $\sim 8.6\pm0.1$ dex, which is consistent with the metallicity obtained from Eq.~\ref{eq:metallicity}. The central region shows higher metallicity, possibly because of star formation triggered by the accumulation of gas funnelled from the outer disk into the inner region by the bar. Using the baryonic Tully-Fisher relation\citep{McGaugh2000ApJ...533L..99M}, the disk rotation velocity is 86$^{+14}_{-11}$ km s$^{-1}$. We found out that the inclination of \newgalaxy\ is around 13$^{\circ}$ using:
\begin{equation}
V_{OBS}= V_{ROT}\times sin(i) 
\end{equation}
where V$_{OBS}$ and V$_{ROT}$ are the observed and actual rotation velocity of galaxy and \textit{i} represents the inclination of the galaxy.

The mass surface density is shown in the panel (d) of Fig.~\ref{fig:brightness_profile}. The inner disk has a stellar mass surface density, $\Sigma(*)\approx$0.5 to 2.5 $\times$10$^{7} M_{\odot}kpc^{-2}$, whereas the outer disk has $\Sigma(*)\leq$4 $\times$10$^{6} M_{\odot}kpc^{-2}$. The inner disk extends out to $\sim$20$^{\prime\prime}$ or four kpc, and the outer disk extends out to $\sim$45$^{\prime\prime}$ or 9.1 kpc. It is not surprising that only the inner disk shows star formation, as the outer disk is too diffuse and probably does not have enough disk gravity for gas to condense and form stars.

\subsection{Star Formation History and Star Formation Rate in UV and H$\alpha$}
\begin{figure}
    \centering
    \includegraphics[width=0.45\textwidth]{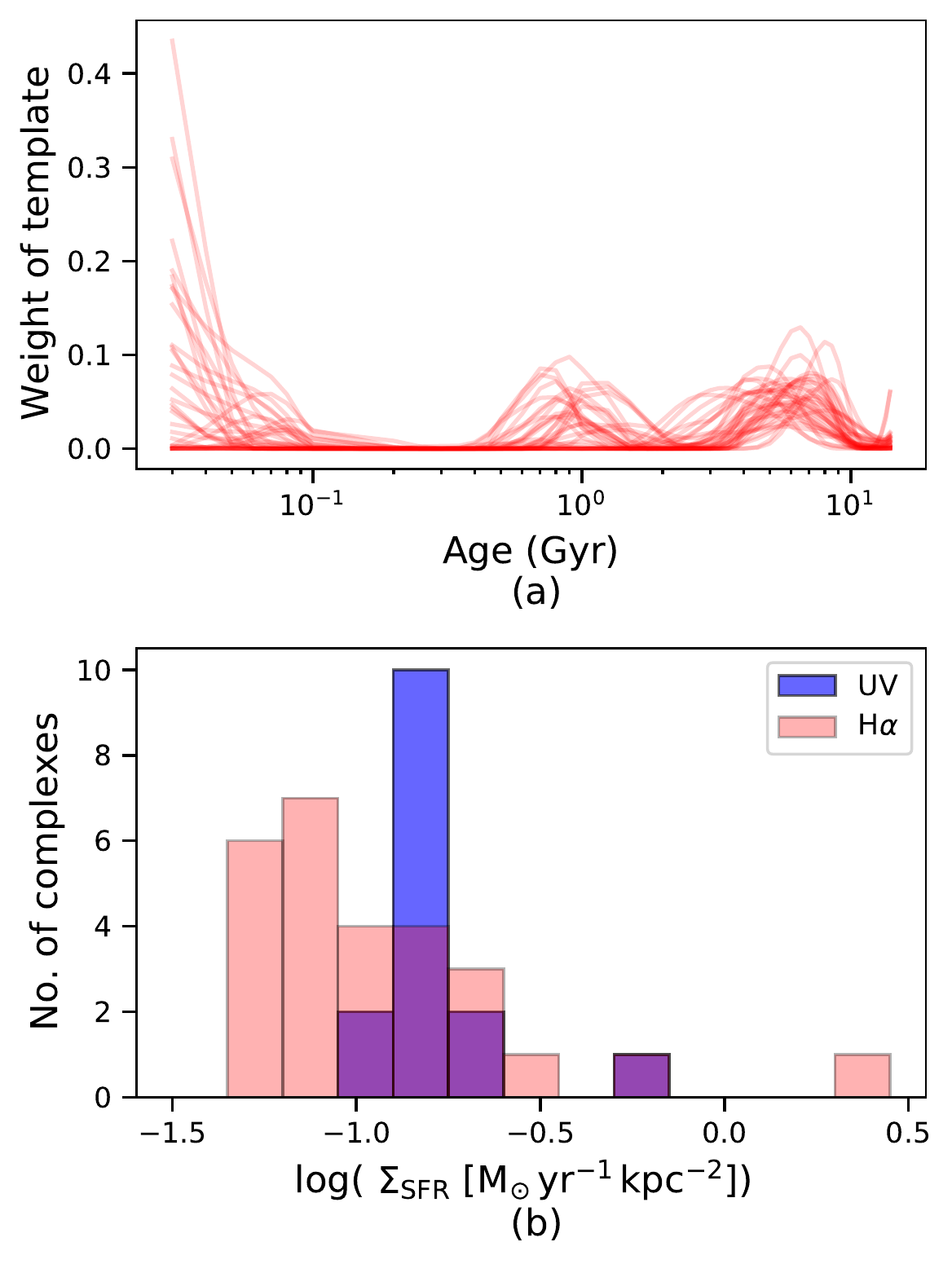}
    \caption{The top panel gives the fraction of the stellar light that originates from a stellar population as a function of age. The curves represent individual bins from sec.~\ref{sec:IFU}. The bottom panel shows the histograms of log($\Sigma_{SFR}$) of identified SFCs in UV and H$\alpha$.}
    \label{fig:SFH_SFR_UV_Ha}
\end{figure}

Fig.~\ref{fig:SFH_SFR_UV_Ha}(a) shows the star formation history of the galaxy derived from \textsc{gist}. The galaxy has gone through three bursts of star formation, separated by a few Gyrs each. The latest episode of star formation is ongoing in this galaxy, which is also evident from the UV and H$\alpha$ images, which reveal the SFCs. We extracted the SFCs from the FUV F148W and H$\alpha$ images using the Python library for Source Extraction and Photometry (\textsc{sextractor}; \citealt{Bertin1996}). We used a detection threshold of 5$\sigma$, where $\sigma$ is the global background noise. We detected 15 SFCs in FUV and 27 SFCs in H$\alpha$, respectively. The number of detected FUV SFCs is lower because the FUV observations are not as deep as the H$\alpha$ observations, so the fainter SFCs in H$\alpha$ may have been missed in FUV. The spatial resolution of the MUSE H$\alpha$ image is also higher than the UVIT FUV image. 

We calculated the star formation rate per unit area ($\Sigma_{SFR}$) for each of the identified SFCs by performing elliptical aperture photometry on the identified sources using \textsc{photutils}, a python \textsc{astropy} package for photometry. We calculated the $\Sigma_{SFR}$ of each SFC from FUV emission using \citet{ Salim2007} (see \citealt{Yadav2021ApJ...914...54Y} for more details) and H$\alpha$ using \citet{Calzetti2007ApJ...666..870C}. Fig.~\ref{fig:SFH_SFR_UV_Ha} (a) and (b) shows the star formation history and $\Sigma_{SFR}$ for FUV and H$\alpha$ of \newgalaxy, respectively. The mean log($\Sigma_{SFR}$(M$_\odot$ yr$^{-1}$ kpc$^{-2}$)) estimated for \newgalaxy\ from FUV and H$\alpha$ is -0.74 and -0.68 which is higher than XUV regions in galaxies as estimated by \citet{Thilker2007}. 

\section{Discussion} \label{sec:discussion}
We present the serendipitous discovery of a star-forming galaxy \newgalaxy\ at a redshift of 0.00980$\pm$0.00018 (D $\approx$ 41.86 Mpc), lying in the foreground of NGC\,6902A (z=0.05554$\pm$0.00010, D=253 Mpc; \citealt{Costa1991ApJS...75..935D}). They were earlier classified as an interacting system \citep{1991deVaucouleurs}. However, the MUSE data clearly shows that both galaxies are at different redshifts and are not interacting with each other.

NGC\,6902A shows signatures of a small perturbation of old stars towards the south-west region in the K band. A few galaxies in the field have similar photometric redshifts, so that the perturbed region could be due to a past dry merger. NGC\,6902A is morphologically classified as SB(s)m pec in RC3 catalogue \citep{deVaucouleurs1991S&T....82Q.621D} and one of the Sersic functions used in our \textsc{galfit} decomposition represents a bar, thus confirming the RC3 morphology.

Our detection of \newgalaxy\ shows that diffuse star-forming galaxies can be missed in surveys but can be detected in UV and H$\alpha$ observations, so the latter are very important for discovering diffuse LSBGs or UDGs at low redshifts. This is similar to the discovery of XUV disks via UV observations \citep{Thilker2007}. Also, spectroscopic observations are very important for avoiding projection bias and for detecting new galaxies \citep{Yadav2021A&A...651L...9Y}. 

The star formation history (Fig.~\ref{fig:SFH_SFR_UV_Ha}) of \newgalaxy\ shows that the galaxy has gone through two bursts of star formation previously, and the ongoing third burst is very prominent in UV and H$\alpha$. Fig.~\ref{fig:SFH_SFR_UV_Ha} bottom panel shows that the star formation rate of the extracted SFCs has similar mean values for FUV and H$\alpha$ emission. \newgalaxy\ also has spiral arms and a bar. Fig.~\ref{fig:vel_metall_sfh} shows that the spiral arm and bar in \newgalaxy\ have enhanced metallicity. There is also higher metallicity in the outer disk than in the inner disk. The metal-enriched gas will increase gas cooling and can lead to further star formation in the central region \citep{Martel2013MNRAS.431.2560M}, contributing to the buildup of the inner disk. The star formation history (Fig.~\ref{fig:SFH_SFR_UV_Ha}) shows that \newgalaxy\ is currently in the phase of rising star formation, so we expect the galaxy to become more luminous. Thus, \newgalaxy\ is an example of a diffuse, LSBG transforming into a luminous galaxy.
 
The central disk brightness for the outer diffuse disk in the r band is {close} to the LSBG cutoff of $\mu_{0}(r)=21$ mag arcsec$^{-2}$ \citep{brown.etal.2001}. Hence, although the inner disk is star-forming, the outer disk is diffuse and similar to an LSBG in nature. The effective radius of \newgalaxy\ {in r band} is {7.57} kpc with a Sersic index of {0.52}, which is shallower than exponential decline (n=1) and similar to spiral galaxies. The $\Sigma(*)$ of the inner disk is similar to the outer disks of XUV galaxies \citep{das.etal.2020}, whereas the $\Sigma(*)$ of the outer disk is similar to dwarf LSBGs. The central surface brightness of \newgalaxy\ is close to the limiting value for LSBGs (see Section 3.2), which are known to be dark matter dominated galaxies \citep{deblok.mcgaugh.1997}. However, determining the dark matter mass of \newgalaxy\ is difficult because the galaxy is close to face-on. So the HI line width or H$\alpha$ velocity field will not give an accurate estimate of the disk rotation velocity. Future deep observations of the HI distribution can perhaps reveal something about the dark matter content.
Thus, we conclude that \newgalaxy\ is a diffuse, gas-rich LSBG in the process of disk growth via star formation in its inner disk. Its detection raises the possibility of finding similar diffuse systems at low redshifts using star formation tracers such as UV and H$\alpha$. It also raises the intriguing question of how star formation is supported in such diffuse disks and whether disk dark matter may be playing a role \citep{das.etal.2020}.

\section{Conclusions}
\label{sec:conclusions}
This letter presents the serendipitous discovery of a star-forming galaxy UVIT\, J202258.73-441623.8 at a redshift of 0.00980$\pm$0.00018 lying in the foreground of NGC\,6902A, which is mistakenly classified as an interacting galaxy. Based on the detailed study using UVIT, DECaLS and MUSE data, the following conclusions are inferred:

\begin{enumerate}
\item \newgalaxy\ shows ongoing star formation in UV and H$\alpha$. It also shows a spiral arm and a bar-like feature.
\item The mean {log($\Sigma_{SFR}$ (M$_\odot$ yr$^{-1}$ kpc$^{-2}$))} of \newgalaxy\ in FUV and H$\alpha$ is -0.74 and -0.68 respectively.
\item \newgalaxy\ has gone through three starburst phases, and the latest episode of star formation is ongoing.
\item The large outer, diffuse disk is low in surface brightness (LSB) and has a low stellar density. So this galaxy is an example of a galaxy with an extended outer low surface brightness disk, which is transforming into a luminous galaxy. 
\end{enumerate}

The discovery of a new foreground galaxy which was mistaken to be a tidal feature of a bright background galaxy, using powerful instruments like UVIT and MUSE thus opens a gateway to search for similar cases, where blue diffuse tidal features in interacting galaxies may not be the remnant of a merger but instead is a separate foreground and/or background galaxy. It also shows the power of using star formation tracers such as FUV and H$\alpha$ emission to detect diffuse galaxies.
    
\begin{acknowledgements}
We thank the anonymous referee for the thoughtful review, which improved the impact and clarity of this work. This paper has used the data from the UVIT, which is part of the AstroSat mission of the Indian Space Research Organization (ISRO), archived at the Indian Space Science Data Center (ISSDC). This publication has also used near-infrared data from IRSF at SAAO. This paper has also used the observations collected at the European Southern Observatory under ESO programme 0103.A-0637, run B. This research has also used data from DECaLS at CTIO. This publication has used the NASA/IPAC Extragalactic Database (NED), which is operated by the Jet Propulsion Laboratory, California Institute of Technology, under contract with the National Aeronautics and Space Administration. MD acknowledges the support of the Science and Engineering Research Board (SERB) MATRICS grant MTR/2020/000266 for this research. 
\end{acknowledgements}

%
  \bibliographystyle{aa} 
%
\bibliography{sample}
\onecolumn

\end{document}